\documentclass[12pt]{article}

\usepackage{graphicx}
\usepackage{here}
\usepackage{cite}
\usepackage{epsfig}
\usepackage{amsfonts}
\usepackage{amsmath}
\usepackage{amssymb}

\newcommand{\sla}[1]{/\!\!\!#1}
\def\lsim{\mathrel{\raise.3ex\hbox{$<$\kern-.75em\lower1ex\hbox{$\sim$}}}}
\def\gsim{\mathrel{\raise.3ex\hbox{$>$\kern-.75em\lower1ex\hbox{$\sim$}}}}

\begin{document}
\thispagestyle{empty}

\def\thefootnote{\fnsymbol{footnote}}

\begin{flushright}

CERN--PH--TH/2004--103 \hfill
DCPT/04/56 \\
DESY--04--089 \hfill
IPPP/04/28 \\
KA--TP--04--2004\hfill
MADPH--04--1384 \\
hep-ph/0406323 \hfill
\end{flushright}

\vspace{1cm}

\begin{center}

{\Large\sc {\bf Extracting Higgs boson couplings from LHC data}}

\vspace{1cm}

{\sc 
M.~D\"uhrssen$^{1}$%
\footnote{email: michael.duehrssen@gmx.de}%
, S.~Heinemeyer$^{2}$%
\footnote{email: Sven.Heinemeyer@cern.ch}%
, H.~Logan$^{3}$%
\footnote{email: logan@pheno.physics.wisc.edu}%
, D.~Rainwater$^{4}$%
\footnote{email: rain@mail.desy.de}, \\[.2em]
G.~Weiglein$^{5}$%
\footnote{email: Georg.Weiglein@durham.ac.uk}%
~and D.~Zeppenfeld$^{3,6}$%
\footnote{email: dieter@particle.uni-karlsruhe.de}
}

\vspace*{1cm}

{\sl
$^1$Physikalisches Institut, Universit\"at Freiburg, Hermann-Herder-Str.\ 3,\\ 
D--79104 Freiburg, Germany

\vspace*{0.3cm}

$^2$CERN TH Division, Dept.\ of Physics,
CH-1211 Geneva 23, Switzerland
 
\vspace*{0.3cm}

$^3$Dept.\ of Physics, University of Wisconsin, 1150 University
Avenue, \\
Madison, WI 53706 USA

\vspace*{0.3cm}

$^4$DESY Theory, Notkestr.\ 85, D--22603 Hamburg, Germany

\vspace*{0.3cm}

$^5$Institute for Particle Physics Phenomenology, University of Durham,\\
Durham DH1~3LE, UK

\vspace*{0.3cm}

$^6$Institut f\"ur Theoretische Physik, Universit\"at Karlsruhe, 
D--76128 Karlsruhe, Germany
}

\end{center}

\vspace*{0.5cm}

\begin{abstract}
  We show how LHC Higgs boson production and decay data can be used to
  extract gauge and fermion couplings of Higgs bosons. We show that
  very mild theoretical assumptions, which are valid in general
  multi-Higgs doublet models, are sufficient to allow the extraction
  of absolute values for the couplings rather than just ratios of the
  couplings.  For Higgs masses below 200~GeV we find accuracies of
  $10-40\%$ for the Higgs boson couplings and total width after
  several years of LHC running.  Slightly stronger assumptions on the
  Higgs gauge couplings even lead to a determination of couplings to
  fermions at the level of $10-20\%$.  We also study the sensitivity
  to deviations from SM predictions in several supersymmetric
  benchmark scenarios as a subset of the analysis.
\end{abstract}

\def\thefootnote{\arabic{footnote}}
\setcounter{page}{0}
\setcounter{footnote}{0}

\newpage

%%%%%%%%%%%%%%%%%%%%%%%%%%%%%%%%%%%%%%%%%%%%%%%%%%%%%%%%%%%%%%%%%%%%%%%%%%%%%%%
%%%%%%%%%%%%%%%%%%%%%%%%%%%%%%%%%%%%%%%%%%%%%%%%%%%%%%%%%%%%%%%%%%%%%%%%%%%%%%%

\section{Introduction}

LHC experiments will have the capability to observe a Higgs boson,
Standard Model (SM) or non, in a variety of channels, in particular if
its mass lies in the intermediate mass region, $114<m_H\lsim 250$~GeV,
as suggested by direct searches~\cite{Barate:2003sz} and electroweak
precision data~\cite{Grunewald:2003ij}.  Once the Higgs boson is
discovered and its mass measured, one will want to gain as much
information as possible on its couplings to both gauge bosons and
fermions.  These measurements will provide crucial tests of the mass
generation mechanism realized in nature.

The various Higgs couplings determine Higgs production cross sections
and decay branching fractions. By measuring the rates of multiple
channels, various combinations of couplings can be determined. A
principal problem at the LHC is that there is no technique analogous
to the measurement of the missing mass spectrum at a linear
collider~\cite{Garcia-Abia:1999kv} which would directly determine the
total Higgs production cross section.  The decay $H\to b\bar{b}$,
which by far dominates for a light Higgs boson, will be detectable but
suffer from large experimental uncertainties.  In addition, some Higgs
decay modes cannot be observed at the LHC.  For example, $H\to gg$ or
decays into light quarks will remain hidden below overwhelming QCD
dijet backgrounds.  This implies that absolute measurements of
(partial) decay widths are possible only with additional theoretical
assumptions.

One possible strategy was outlined in
Refs.~\cite{Djouadi:2000gu,Zeppenfeld:2000td,Belyaev:2002ua}.
Assuming the absence of unexpected decay channels and a SM ratio of
the $H\to b\bar{b}$ and $H\to\tau\tau$ partial widths, absolute
measurements of $\Gamma(H\to WW,ZZ)$, $\Gamma(H\to\tau\tau)$,
$\Gamma(H\to\gamma\gamma)$, $\Gamma(H\to gg)$ and of the top quark
Yukawa coupling squared, $Y_t^2$, are possible, with errors in the
$10-30\%$ range.

Here we revisit the information that could be extracted at the LHC
from rate measurements of an intermediate-mass Higgs boson.  We
consider the expected accuracies at various stages of the LHC program:
after 30~fb$^{-1}$ of low luminosity ($10^{33}\,{\rm cm}^{-2}{\rm
  sec}^{-1}$) running, 300~fb$^{-1}$ at high luminosity
($10^{34}\,{\rm cm}^{-2}{\rm sec}^{-1}$), and a mixed scenario where
the weak boson fusion channels are assumed to suffer substantially
from pile-up problems under high luminosity running conditions (making
forward jet tagging and central jet veto fairly inefficient).

A rather model-independent analysis, where only ratios of couplings
(or partial widths) could be extracted, was performed in
Ref.~\cite{ATL-PHYS-2003-030}. Here we consider general
multi-Higgs-doublet models (with or without additional Higgs
singlets), in which the $HWW$ and $HZZ$ couplings are bounded from
above by their SM values, i.e., we impose theoretically motivated
constraints on these two couplings. These constraints are valid, in
particular, for the Minimal Supersymmetric Standard Model (MSSM), and
will sharpen the implications of LHC data for Higgs couplings very
significantly.

As an illustration of coupling measurements at the LHC, we consider
specific MSSM benchmark scenarios of Ref.~\cite{Carena:2002qg} as
examples.  The significance of deviations of the measured rates from
SM predictions provides a measure of the sensitivity of LHC
measurements in the Higgs sector.

The paper is organized as follows. In Section~\ref{sec:channels} we
review the existing analyses on the production and decay channels used
throughout our paper, followed by our theoretical assumptions and
fitting procedure details in Sections~\ref{sec:nomodel}
and~\ref{sec:mssm_specific}, respectively for general
multi-Higgs-doublet models and a specific MSSM scenario.
Section~\ref{sec:sum_out} contains a summary and outlook.

%%%%%%%%%%%%%%%%%%%%%%%%%%%%%%%%%%%%%%%%%%%%%%%%%%%%%%%%%%%%%%%%%%%%%%%%%%%%%%%
%%%%%%%%%%%%%%%%%%%%%%%%%%%%%%%%%%%%%%%%%%%%%%%%%%%%%%%%%%%%%%%%%%%%%%%%%%%%%%%

\section{Summary of Higgs boson channels}
\label{sec:channels}

In order to determine the properties of a physical state such as a
Higgs boson, one needs at least as many separate measurements as
properties to be measured, although two or more measurements can be
made from the same channel if different information is used, e.g.,
total rate and an angular distribution.  Fortunately, the LHC will
provide us with many different Higgs observation channels.  In the SM
there are four relevant production modes: gluon fusion (GF;
loop-mediated, dominated by the top quark), which dominates inclusive
production; weak boson fusion (WBF), which has an additional pair of
hard and far-forward/backward jets in the final state; top-quark
associated production ($t\bar{t}H$); and weak boson associated
production ($WH,ZH$), where the weak boson is identified by its
leptonic decay.~\footnote{We do not consider diffractive Higgs
  production since its rate is in general small and also quite
  uncertain, which limits the usefulness of this channel for Higgs
  coupling determinations.}

Although a Higgs is expected to couple to all SM particles, not all
these decays would be observable.  Very rare decays (e.g., to
electrons) would have no observable rate, and other modes are
unidentifiable QCD final states in a hadron collider environment
(gluons or quarks lighter than bottom).  In general, however, the LHC
will be able to observe Higgs decays to photons, weak bosons, tau
leptons and $b$ quarks, in the range of Higgs masses where the
branching ratio (BR) in question is not too small.

For a Higgs in the intermediate mass range, the total width, $\Gamma$,
is expected to be small enough to use the narrow-width approximation
in extracting couplings.  The rate of any channel (with the $H$
decaying to final state particles $xx$) is, to good approximation,
given by
\begin{equation}
\sigma(H)\times {\rm BR}(H\to xx) = {\sigma(H)^{\rm SM}\over \Gamma_p^{\rm SM}}
\cdot {\Gamma_p\Gamma_x \over \Gamma}\;,
\end{equation}
where $\Gamma_p$ is the Higgs partial width involving the production
couplings and where the Higgs branching ratio for the decay is written
as ${\rm BR}(H\to xx)=\Gamma_x/\Gamma$. Even with cuts, the observed
rate directly determines the product $\Gamma_p\Gamma_x/\Gamma$
(normalized to the calculable SM value of this product).  The LHC will
have access to (or provide upper limits on) combinations of
$\Gamma_g,\Gamma_W,\Gamma_Z, \Gamma_\gamma,\Gamma_\tau,\Gamma_b$ and
the square of the top Yukawa coupling, $Y_t$.~\footnote{We do not
  write this as a partial width, $\Gamma_t$, because, for a light
  Higgs, the decay $H\to t\bar{t}$ is kinematically forbidden.}

Since experimental analyses are driven by the final state observed, we
classify Higgs channels by decay rather than production mode, and then
discuss the different production characteristics as variants of the
final state.  However, some initial comments on production modes are
in order.  First, experimental studies mostly do not yet include the
very large (N)NLO enhancements known for $gg\to
H$~\cite{Harlander:2002wh,Anastasiou:2002yz,Ravindran:2003um}.  Even
if background corrections are as large as for the signal, which they
typically are not, the statistical significance of the GF channels
will be greater than estimated by the current studies (which we have
used for this paper).  Furthermore, the NNLO calculations may reduce
also the theory systematic uncertainty for the signal.  Second,
experimental studies do not consider WBF channels above 30~fb$^{-1}$
integrated luminosity, because the efficiency to tag forward jets at
high-luminosity LHC running is not yet fully understood.  This is a
very conservative assumption.  We also discuss a scenario where a
higher luminosity is available in the WBF channels.

The literature on Higgs channels at LHC is extensive.  We refer here
only those analyses which we use in our fits.  Mostly, these are
recent experimental analyses which contain references to the earlier
phenomenological proposals.  We always use the rates as summarized in
Ref.~\cite{ATL-PHYS-2003-030} for our fits.
%----------------START CHANNELS
The individual channels used for the fits are:
\begin{itemize}
\item GF  $gg\to H\to ZZ$ \cite{:1999fr,ATL-PHYS-2003-030}

\item WBF $qq\, H\to qq\, ZZ$ \cite{ATL-PHYS-2003-030}

\item GF $gg\to H\to WW$ \cite{:1999fr,ATL-PHYS-2003-030}

\item WBF $qq\, H\to qq\, WW$ \cite{ATL-PHYS-2003-005}

\item $W\, H\to W\, WW$ (2$l$ and 3$l$ final state) \cite{ATL-PHYS-2000-008,ATL-PHYS-2000-013}

\item $t\bar{t}\, H (H\to WW, t\to Wb)$ (2$l$ and 3$l$ final state) \cite{ATL-PHYS-2002-019}

\item Inclusive Higgs boson production: $H\to\gamma\gamma$ \cite{:1999fr}

\item WBF $qq\, H\to qq\,\gamma\gamma$ \cite{ATL-PHYS-2003-036}

\item $t\bar{t}\, H (H\to\gamma\gamma)$ \cite{Eynard:98}

\item $W\, H (H\to\gamma\gamma)$ \cite{Eynard:98}

\item $Z\, H (H\to\gamma\gamma)$ \cite{Eynard:98}

\item WBF $qq\, H\to qq\,\tau\tau$ \cite{ATL-PHYS-2003-005,ATL-COM-PHYS-2003-006,ATL-COM-PHYS-2002-045}

\item $t\bar{t}\, H (H\to b\bar{b})$ \cite{ATL-PHYS-2003-024}
\end{itemize}
%----------------END CHANNELS

For the WBF channels we include a minijet veto, even if the study
cited did not.  In the discussion below, statements about Higgs rates
typically refer to the SM-like case.  Substantially suppressed
branching ratios are possible beyond the SM and may change a
measurement into an upper bound.

%%%%%%%%%%%%%%%%%%%%%%%%%%%%%%%%%%%%%%%%%%%%%%%%%%%%%%%%%%%%%%%%%%%%%%%%%%%%%%%
%%%%%%%%%%%%%%%%%%%%%%%%%%%%%%%%%%%%%%%%%%%%%%%%%%%%%%%%%%%%%%%%%%%%%%%%%%%%%%%

\subsection{$H\to Z^{(*)}Z^{(*)}\to 4\ell$}

Leptons are the objects most easily identified in the final state, so
this decay is regarded as ``golden'' due to its extreme cleanliness
and very low background.  It is a rare decay due to the subdominance
of $H\to ZZ$ relative to $H\to W^+W^-$, and because of the very small
BR of $Z\to\ell^+\ell^-$.  Fortunately, due to the possible decay to
off-shell $Z$~bosons, a SM Higgs has non-negligible BR to $4\ell$ even
for $m_H<2M_Z$, down to approximately 120~GeV.~\footnote{We note that
  for such low masses, doubly off-shell effects must be taken into
  account.}  Due to the low event rate, current studies concentrate on
inclusive measurements which are dominated by GF. They provide
information mainly on the product $\Gamma_g\Gamma_Z/\Gamma$.

The most advanced analysis for this channel~\cite{Cranmer:2003kf} was
made recently by ATLAS. (For an older CMS study,
see~\cite{Puljak.thesis}).  Its principal improvement over previous
studies is the use of all available NLO results (the only study so far
to do this) for both the dominant GF signal and its major backgrounds.
Further improvements can be expected in the inclusion of off-shell
contributions to the $gg\to ZZ$ background, for which ATLAS used an
approximate K-factor.

By isolating the WBF contribution one obtains some independent
information on the product $\Gamma_{W,Z}\Gamma_Z/\Gamma$, in particular if
high-luminosity running can be exploited for this channel.

%%%%%%%%%%%%%%%%%%%%%%%%%%%%%%%%%%%%%%%%%%%%%%%%%%%%%%%%%%%%%%%%%%%%%%%%%%%%%%%
%%%%%%%%%%%%%%%%%%%%%%%%%%%%%%%%%%%%%%%%%%%%%%%%%%%%%%%%%%%%%%%%%%%%%%%%%%%%%%%

\subsection{$H\to\gamma\gamma$}

Photons are also readily identifiable, but are more difficult than
leptons to measure because of a large, non-trivial background from
jets faking photons.  Higgs photonic decay is loop-induced and
therefore rare, even more so because of destructive interference
between the top-quark and $W$ loops.  This is in some sense
advantageous, because this decay mode is thus sensitive to variations
in the weak gauge and top Yukawa couplings and additional particles in
the loop.  This decay is visible in the SM only for the lower Higgs
mass range, $110<m_H<150$~GeV.

Despite the difficulties of identifying photons, which are not yet
fully understood for the LHC, especially for high-luminosity running,
Higgs decays to photons should be observable in both
GF~\cite{:1999fr,:1997kj,Abdullin:1998er} and
WBF~\cite{ATL-PHYS-2003-036,CMS-NOTE-2001/022}, unless ${\rm
  BR}(H\to\gamma\gamma)$ is substantially smaller than in the SM.
These channels measure the products $\Gamma_g\Gamma_\gamma/\Gamma$ and
$\Gamma_{W,Z}\Gamma_\gamma/\Gamma$. The $H\to\gamma\gamma$ signals in
$t\bar{t}H,WH$ and $ZH$ production~\cite{:1999fr,Dubinin:1997rn} are
very weak, due to the paucity of events even at high-luminosity
running, but could be used as supplemental channels, and would be
especially useful if LHC observes a non-SM Higgs.

%%%%%%%%%%%%%%%%%%%%%%%%%%%%%%%%%%%%%%%%%%%%%%%%%%%%%%%%%%%%%%%%%%%%%%%%%%%%%%%
%%%%%%%%%%%%%%%%%%%%%%%%%%%%%%%%%%%%%%%%%%%%%%%%%%%%%%%%%%%%%%%%%%%%%%%%%%%%%%%

\subsection{$H\to W^{+(*)}W^{-(*)}\to\ell^+\ell^- +\sla{p}_T$}

This decay can be observed in
GF~\cite{:1999fr,Dittmar:1997ss,Green:2000um} and
WBF~\cite{ATL-PHYS-2003-005,CMS-NOTE-2001/016} using
$W^+W^-\to\ell^+\ell^- +\sla{p}_T$ final states, as well as in
$t\bar{t}H$ associated production using combinations of multilepton
final states~\cite{ATL-PHYS-2002-019}.  The first two modes extract
the products $\Gamma_g\Gamma_W/\Gamma$ and $\Gamma^2_W/\Gamma$ and are
extremely powerful statistically, while the $t\bar{t}H$ mode can
extract the top Yukawa coupling with high luminosity once $\Gamma_W$
is known.  All these channels are accessible over a wide range of
Higgs masses, approximately $120<m_H<200$~GeV.  An additional
study~\cite{ATL-PHYS-2000-013} of the $WH,H\to WW$ channel for
$m_H>150$~GeV found only a very weak signal, less than $5\sigma$ even
for 300~fb$^{-1}$ of data. We include this channel in our analysis.

The GF mode should improve after (N)NLO effects are included, although
the backgrounds considered did not include off-shell $gg\to
W^{(*)}W^*$.  Also, the single-top background was conservatively
overestimated.  A reanalysis of this channel with updated simulation
tools would be in order.

%%%%%%%%%%%%%%%%%%%%%%%%%%%%%%%%%%%%%%%%%%%%%%%%%%%%%%%%%%%%%%%%%%%%%%%%%%%%%%%
%%%%%%%%%%%%%%%%%%%%%%%%%%%%%%%%%%%%%%%%%%%%%%%%%%%%%%%%%%%%%%%%%%%%%%%%%%%%%%%

\subsection{$H\to\tau^+\tau^-$}

Observing Higgs decays to taus is not possible in GF because of
serious background problems and because the invariant mass of a tau
pair can be reconstructed only when the taus do not decay
back-to-back, which leaves only a small fraction of GF events with
sizable Higgs transverse momentum.  Observation of $H\to\tau^+\tau^-$
is possible in WBF, however~\cite{ATL-PHYS-2003-005,Azuelos:2001yw},
for Higgs masses below about 150~GeV.  As the average Higgs $p_T$ in
this production mode is ${\cal O}(100)$~GeV, the taus are only rarely
produced back-to-back.  This is a relatively rare decay mode, since
BR($H\to\tau\tau$) is typically $5-10\%$ in this mass region and the
taus decay further. At least one tau must decay leptonically, giving
another small BR.  Fortunately, the QCD background to taus is small,
due to excellent fake jet rejection.  While not a discovery channel,
this channel is statistically quite powerful even with only moderate
luminosity, and thus becomes one of the more important decay modes in
a couplings analysis.  This channel measures the product
$\Gamma_{W,Z}\Gamma_\tau/\Gamma$.

%%%%%%%%%%%%%%%%%%%%%%%%%%%%%%%%%%%%%%%%%%%%%%%%%%%%%%%%%%%%%%%%%%%%%%%%%%%%%%%
%%%%%%%%%%%%%%%%%%%%%%%%%%%%%%%%%%%%%%%%%%%%%%%%%%%%%%%%%%%%%%%%%%%%%%%%%%%%%%%

\subsection{$H\to b\bar{b}$}

Associated Higgs-$b$ quark production has too small a cross section in
a SM-like Higgs sector to be observable, so the decay $H\to b\bar{b}$
is the only experimental access to the $b$ Yukawa coupling.  Because
this mode dominates Higgs decays at low mass ($m_H\lsim 135$~GeV
within the SM), an accurate measurement of the bottom Yukawa coupling
is extremely important.  Unfortunately, due to the typically large QCD
backgrounds for $b$ jets, it is very difficult to observe this decay.
The production modes
$t\bar{t}H$~\cite{ATL-PHYS-2003-024,CMS-NOTE-2001/054,Green:2001gh}
and $WH$~\cite{:1999fr,CMS-NOTE-2002/006} might allow very rough
measurements for such a light Higgs, but the statistical significances
are quite low and the background uncertainties quite large and their
rates probably underestimated; they are definitely high-luminosity
measurements.

The $t\bar{t}H$ channel measures the product $Y_t^2\Gamma_b/\Gamma$,
and so would require a separate, precise measurement of $Y_t$ to
isolate $\Gamma_b$.  For $WH$ production, the rate is proportional to
$\Gamma_W\Gamma_b/\Gamma$.  But here the $Wb\bar{b}$ continuum
background has hitherto been underestimated since the NLO QCD
corrections are very large and positive~\cite{Campbell:2003hd}.  A
veto on additional jets may help but requires another detector-level
simulation; unfortunately, it would also increase the background
uncertainty because additional jet activity has been calculated at LO
only.  We include the $t\bar{t}H$ channel but not $WH$ in our
analysis.

%%%%%%%%%%%%%%%%%%%%%%%%%%%%%%%%%%%%%%%%%%%%%%%%%%%%%%%%%%%%%%%%%%%%%%%%%%%%%%%
%%%%%%%%%%%%%%%%%%%%%%%%%%%%%%%%%%%%%%%%%%%%%%%%%%%%%%%%%%%%%%%%%%%%%%%%%%%%%%%

\subsection{Other channels}

The production and decay channels discussed above refer to a single
Higgs resonance, with decay signatures which also exist in the SM. The
Higgs sector may be much richer, of course.  The MSSM with its two
Higgs doublets predicts the existence of three neutral and one charged
pair of Higgs bosons, and the LHC may be able to directly observe
several of these resonances.  Within SUSY models, additional decays,
e.g., into very light super-partners, may be kinematically allowed.
The additional observation of super-partners or of heavier Higgs
bosons will strongly focus the theoretical framework and restrict the
parameter space of a Higgs couplings analysis~\cite{schumitalkPrag}.

At the present time, even enumerating the possibilities is an
open-ended task.  For our present analysis we therefore ignore the
information which would be supplied by the observation of additional
new particles. Instead we ask the better-defined question of how well
LHC measurements of the above decay modes of a single Higgs resonance
can determine the various Higgs boson couplings or partial widths.

%%%%%%%%%%%%%%%%%%%%%%%%%%%%%%%%%%%%%%%%%%%%%%%%%%%%%%%%%%%%%%%%%%%%%%%%%%%%%%%
%%%%%%%%%%%%%%%%%%%%%%%%%%%%%%%%%%%%%%%%%%%%%%%%%%%%%%%%%%%%%%%%%%%%%%%%%%%%%%%

\section{Model assumptions and fits}
\label{sec:nomodel}

In spite of the many decay channels discussed above, the LHC is faced
with the challenge that not all Higgs decay modes can be detected
(e.g., $H\to gg$ is deemed unobservable) or that some important decay
rates, in particular $H\to b\bar{b}$, will suffer from large
experimental uncertainties. In a model-independent analysis, the
limited information which will be available then will lead to strong
correlations in the measurement of different Higgs couplings. These
correlations mask the true precision of LHC measurements when the
expected errors of particular observables like individual partial
widths or branching ratios are considered.

The parameter correlations can be overcome by imposing theoretical
constraints.  One possible approach was suggested in
Refs.~\cite{Djouadi:2000gu,Zeppenfeld:2000td}: fixing the ratio
$\Gamma_b/\Gamma_\tau$ to its SM value, the $H\to\tau\tau$
measurements can be used to pin down the poorly measured Higgs
coupling to bottom quarks.  Here we follow a different approach.  We
perform general fits to the Higgs couplings with the mildest possible
theoretical assumptions, starting with the constraint
\begin{equation}
\Gamma_V\leq\Gamma_V^{\rm SM}
\end{equation}
($V=W,Z$) which is justified in any
model with an arbitrary number of Higgs doublets (with or without
additional Higgs singlets).  I.e., it is true for the MSSM in
particular.

Even without this constraint, the mere observation of Higgs production
puts a lower bound on the production couplings and, thereby, on the
total Higgs width. The constraint $\Gamma_V\leq\Gamma_V^{\rm SM}$,
combined with a measurement of $\Gamma_V^2/\Gamma$ from observation of
$H\to VV$ in WBF, then puts an upper bound on the Higgs total width,
$\Gamma$. It is this interplay which provides powerful constraints on
the remaining Higgs couplings, allowing for their absolute
determination, rather than simply ratios of their magnitudes.

%%%%%%%%%%%%%%%%%%%%%%%%%%%%%%%%%%%%%%%%%%%%%%%%%%%%%%%%%%%%%%%%%%%%%%%%%%%%%%%
%%%%%%%%%%%%%%%%%%%%%%%%%%%%%%%%%%%%%%%%%%%%%%%%%%%%%%%%%%%%%%%%%%%%%%%%%%%%%%%

\subsection{Fitting procedure}
\label{subsec:fitproc}

Our analysis of expected LHC accuracies closely follows the work of
D\"uhrssen~\cite{ATL-PHYS-2003-030}.  First, a parameter space (${\bf
  x}$) is formed of Higgs couplings together with additional partial
widths to allow for undetected Higgs decays and additional
contributions to the loop-induced Higgs couplings to photon pairs or
gluon pairs due to non-SM particles running in the loops.  We assume
that the measured values correspond to the SM expectations for the
purpose of determining statistical uncertainties, then form a log
likelihood function, $L({\bf x})$, which, for a given integrated
luminosity, is based on the expected Poisson errors of the channels
listed in Sec.~\ref{sec:channels} and on estimated systematic
errors~\cite{ATL-PHYS-2003-030}, which are tabulated in the Appendix.

As an alternative, in particular for the specific MSSM scenarios
discussed in Sec.~\ref{sec:mssm_specific}, we use a Gaussian
approximation to the log likelihood function, i.e., a $\chi^2$
function constructed from the same error assumptions that enter the
log likelihood function.  We take each of the channels considered in
Ref.~\cite{ATL-PHYS-2003-030} as a bin in the $\chi^2$. To mimic the
effect of Poisson statistics on channels with low numbers of events,
we discard any channel with $\leq 5$ total events (signal plus
background) in both approaches.  This is relevant only in the case of
low luminosity data.  We have checked that the resulting accuracy
estimates for coupling measurements are consistent for the two
approaches.

Relative to SM expectations, we compute the variation of either
$2L({\bf x})$ or $\chi^2({\bf x})$ on this parameter space and trace
out the surface of variations by one unit.  The $1\sigma$
uncertainties on each parameter are determined by finding the maximum
deviation of that parameter from its SM value that lies on the $\Delta
\chi^2 = 1$ ($\Delta L = 1/2$) surface.  We repeat the procedure for
each Higgs mass value in the range $110\leq m_H\leq 190$ GeV in steps
of 10 GeV.

We perform the fits under three luminosity assumptions for the LHC:
\begin{enumerate}
\item %``low luminosity'': 
  30 fb$^{-1}$ at each of two experiments, denoted
  \underline{2$\times$30 fb$^{-1}$};
\item %``medium luminosity'':
  300 fb$^{-1}$ at each of two experiments, of which only 100
  fb$^{-1}$ is usable for WBF channels at each experiment, denoted
  \underline{2$\times$300 + 2$\times$100 fb$^{-1}$};
\item %``high luminosity'':
  300 fb$^{-1}$ at each of two experiments, with the full luminosity
  usable for WBF channels, denoted \underline{2$\times$300 fb$^{-1}$}.
\end{enumerate}
The second case allows for possible significant degradation of the WBF
channels in a high luminosity environment, while the third case shows
the benefits of additional improvements in WBF studies at high
luminosity.

In both cases the Higgs boson mass is not fitted, i.e. it is assumed
that the mass of the Higgs boson can be measured with high precision
($\Delta m_H/m_H<1\%$) in $H\to Z^{(*)}Z^{(*)}\to 4\ell$ or
$H\to\gamma\gamma$. If both of these channels go unobserved, the
theoretical calculations of Higgs boson BRs get a large error due to
the relatively low precision and larger systematic errors of mass
measurements in WBF $H\to\tau\tau$ or $H\to WW$. This is in itself not
a problem for the measurement of products of Higgs boson couplings,
but a comparison to theoretical prediction, i.e. the upper limit on
$\Gamma_V$, will suffer considerably from this additional uncertainty.

%%%%%%%%%%%%%%%%%%%%%%%%%%%%%%%%%%%%%%%%%%%%%%%%%%%%%%%%%%%%%%%%%%%%%%%%%%%%%%%
%%%%%%%%%%%%%%%%%%%%%%%%%%%%%%%%%%%%%%%%%%%%%%%%%%%%%%%%%%%%%%%%%%%%%%%%%%%%%%%

\subsection{General multi-Higgs-doublet model fits}

We begin by fitting for the uncertainties in the Higgs
couplings-squared in the most general scenario that we consider: we
assume only that
\begin{eqnarray}
g^2(H,W)<1.05 \cdot g^2(H,W,{\rm SM}) \\\nonumber
g^2(H,Z)<1.05 \cdot g^2(H,Z,{\rm SM}) \; .
\end{eqnarray}
Any model that contains only Higgs doublets and singlets will satisfy
the relations without the factor 1.05; the extra $5\%$ margin allows
for theoretical uncertainties in the translation between
couplings-squared and partial widths, and also for small admixtures of
exotic Higgs states, like SU(2) triplets.  We allow for the
possibility of additional particles running in the loops for
$H\to\gamma\gamma$ and $gg\to H$, fitted by a positive or negative new
partial width to these contributions.  Additional light hadronic
decays of the Higgs boson are fitted with a partial width for
undetected decays.  (Invisible decays, e.g.\ to neutralinos could
still be observable~\cite{Eboli:2000ze}.)

%%%%%%%%%%%%%%%%%%%%%%%%% F I G U R E %%%%%%%%%%%%%%%%%%%%%%%%%%%%%%%%%%%%%%%%%
\begin{figure}[thb]
\begin{center}
\resizebox{\textwidth}{!}{
\includegraphics{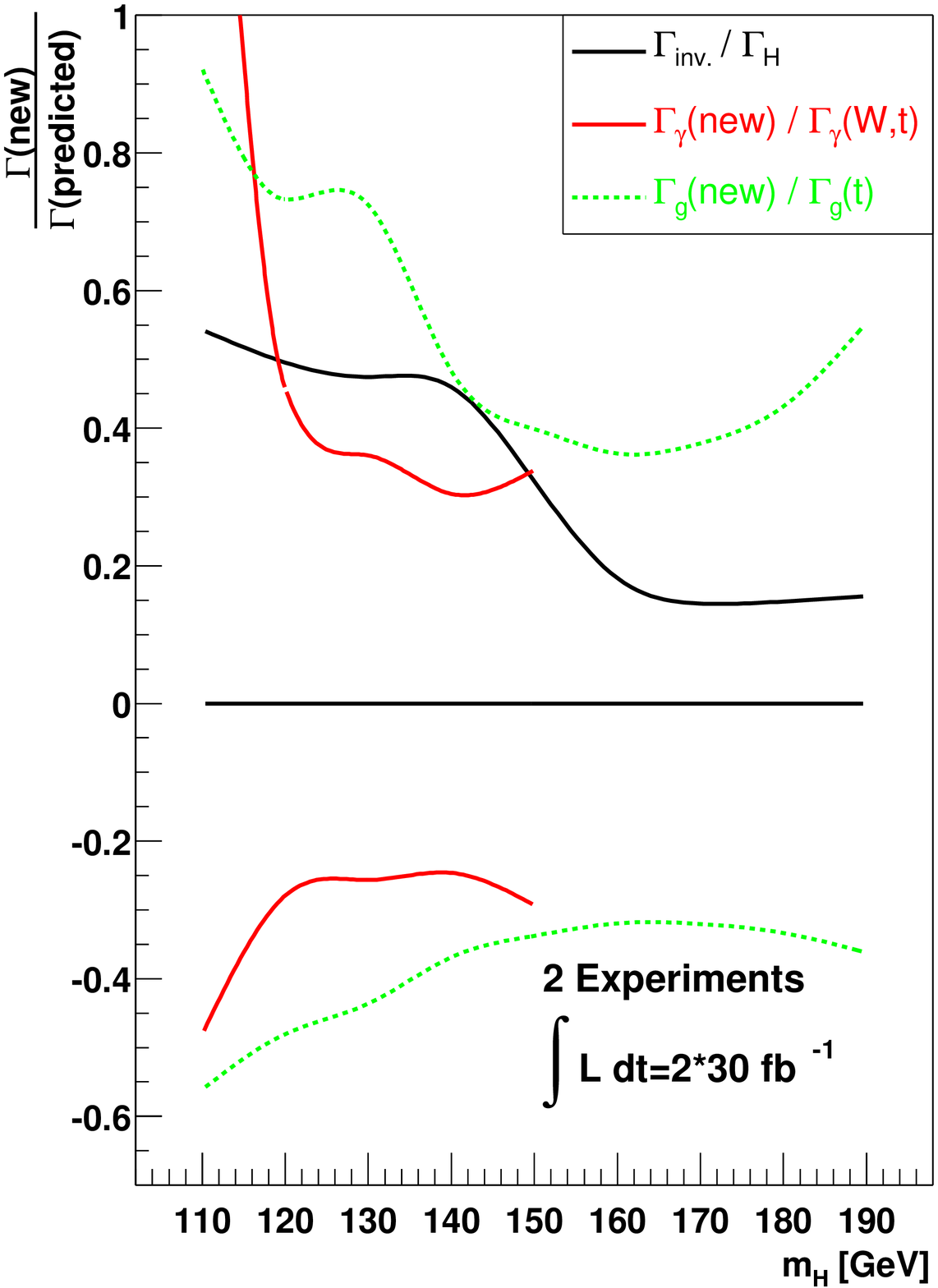}
\includegraphics{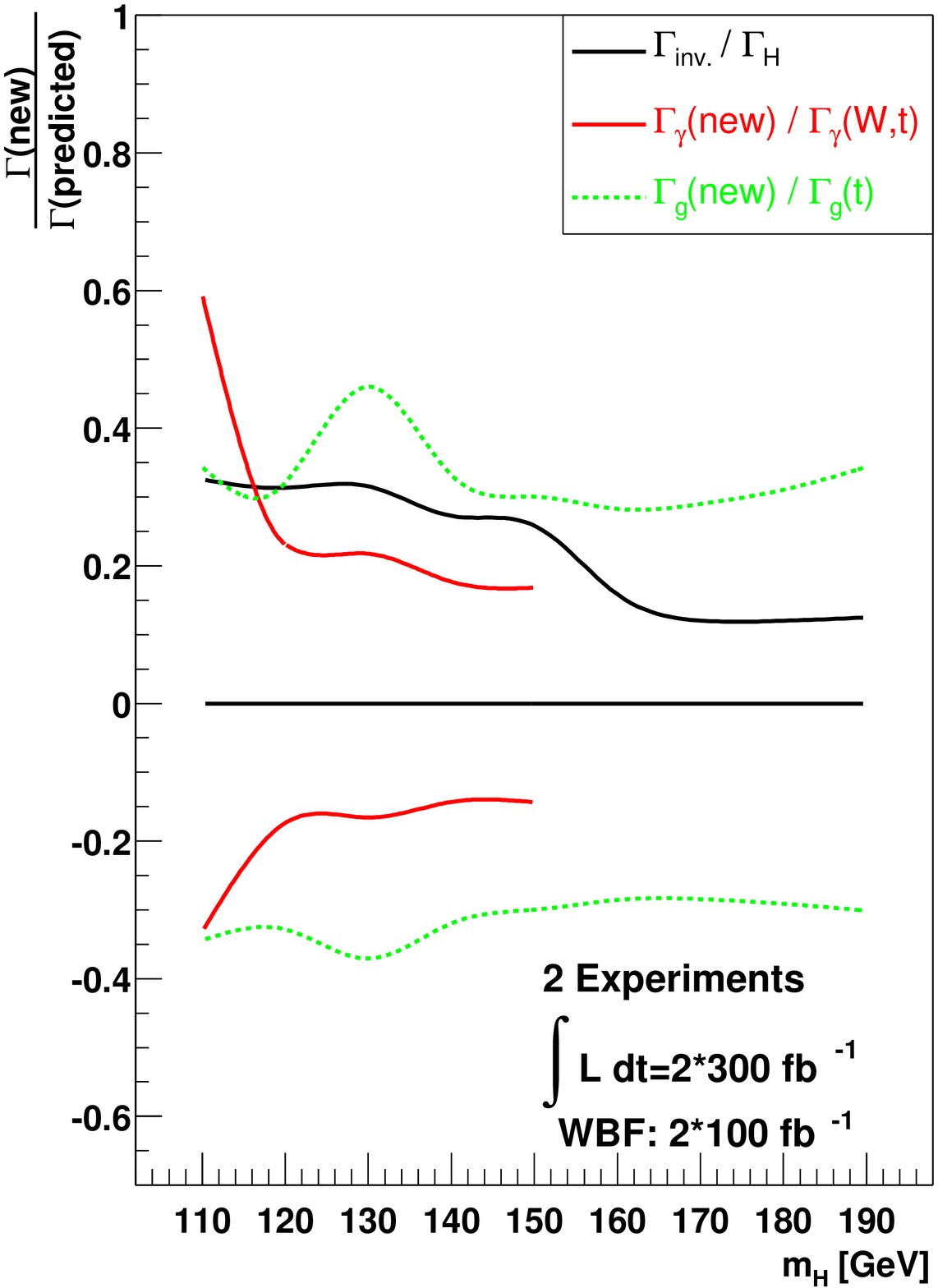}
}
\caption{Relative precisions of fitted new partial widths as a function
  of the Higgs boson mass assuming SM rates and 30~fb$^{-1}$ at each
  of two experiments (left) and 300~fb$^{-1}$ at each of two
  experiments for all channels except WBF, for which 100 fb$^{-1}$ is
  assumed (right).  The new partial width can be due to new particles
  in the loops for $H\to\gamma\gamma$ and $gg\to H$ or due to
  unobservable decay modes.  See text for details.  Here we make the
  weak assumption that $g^2(H,V)<1.05 \cdot g^2(H,V,{\rm SM})$
  ($V=W,Z$).}
\label{fig:Gnew}
\end{center}
\end{figure}
%%%%%%%%%%%%%%%%%%%%%%%%% F I G U R E %%%%%%%%%%%%%%%%%%%%%%%%%%%%%%%%%%%%%%%%%

%%%%%%%%%%%%%%%%%%%%%%%%% F I G U R E %%%%%%%%%%%%%%%%%%%%%%%%%%%%%%%%%%%%%%%%%
\begin{figure}[thb]
\begin{center}
\resizebox{\textwidth}{!}{
\includegraphics{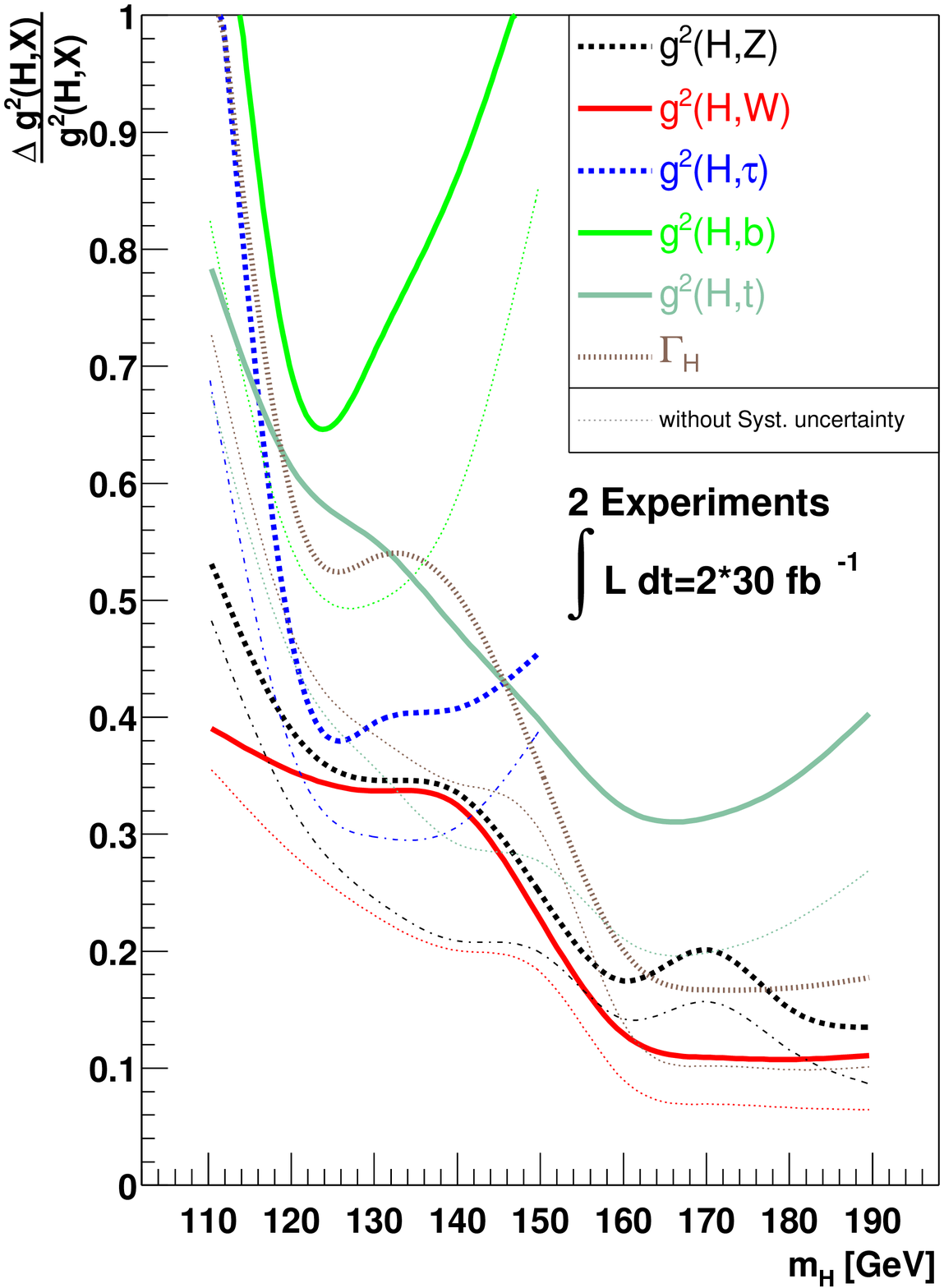}
\includegraphics{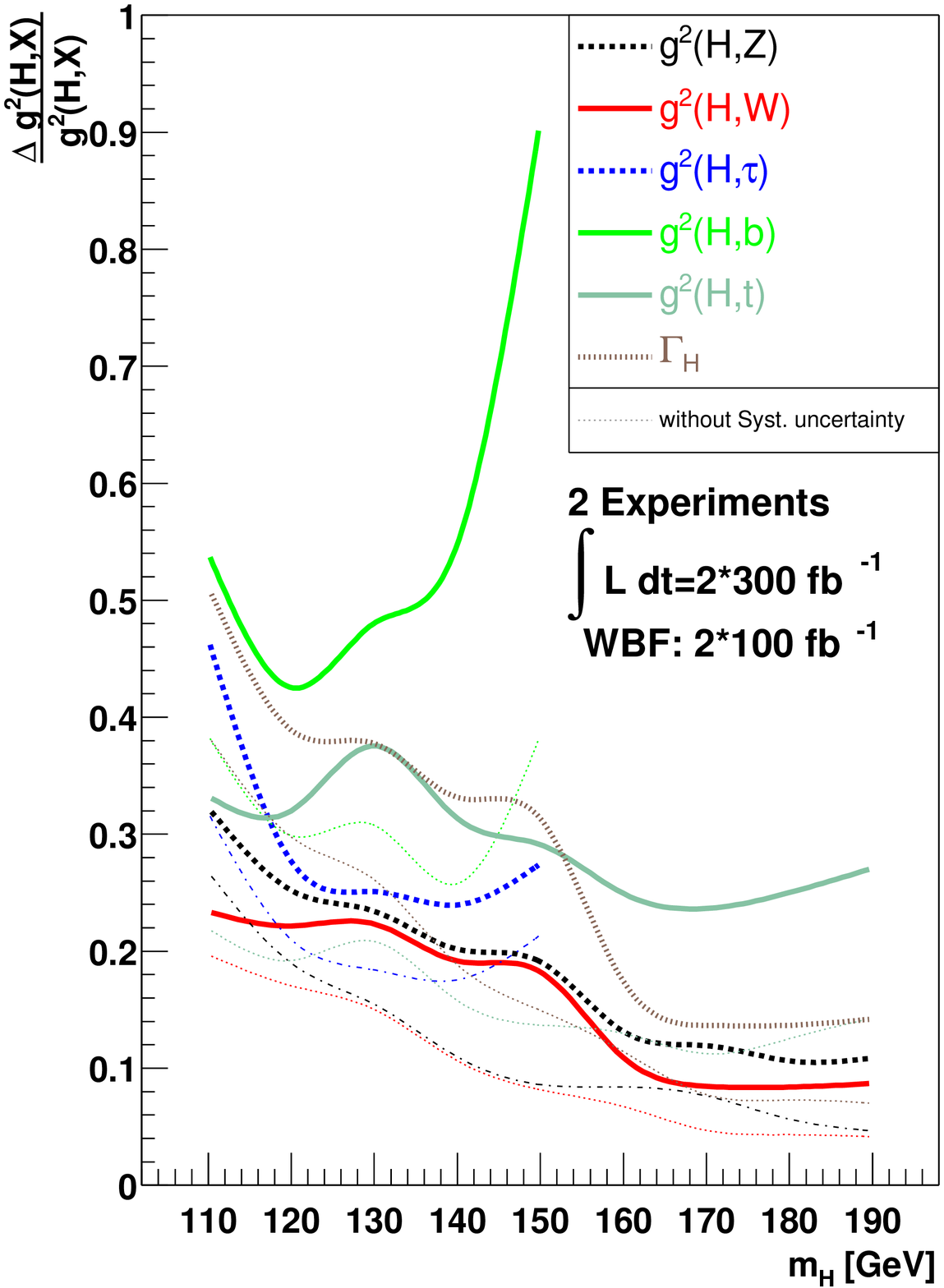}
}
\caption{Relative precision of fitted Higgs couplings-squared as a function
  of the Higgs boson mass for the 2$\times$30~fb$^{-1}$ (left) and the
  2$\times$300 + $2\times$100~fb$^{-1}$ (right) luminosity scenarios
  for SM rates.  Here we make the weak assumption that $g^2(H,V)<1.05
  \cdot g^2(H,V,{\rm SM})$ ($V=W,Z$) but allow for new particles in
  the loops for $H\to\gamma\gamma$ and $gg\to H$ and for unobservable
  decay modes. See text for details.}
\label{fig:fit}
\end{center}
\end{figure}
%%%%%%%%%%%%%%%%%%%%%%%%% F I G U R E %%%%%%%%%%%%%%%%%%%%%%%%%%%%%%%%%%%%%%%%%

The results for the constraints on the new partial widths are shown in
Fig.~\ref{fig:Gnew} as a function of Higgs mass for the
2$\times$30~fb$^{-1}$ and 2$\times$300 + 2$\times$100~fb$^{-1}$
luminosity scenarios and SM rates observed.  The new partial width for
$H\to\gamma\gamma$ is most tightly constrained for $120\lsim m_H\lsim
140$ GeV, being less than $\pm (25-35)\%$ of $\Gamma_\gamma^{\rm SM}$
for 2$\times$30~fb$^{-1}$ and $\pm (10-15)\%$ for 2$\times$300 +
2$\times$100~fb$^{-1}$.  The new partial width for $gg\to H$ is less
well constrained, being less than $\pm (30-90)\%$ of $\Gamma_g^{\rm
  SM}$ for 2$\times$30~fb$^{-1}$ and $\pm (30-45)\%$ for 2$\times$300
+ 2$\times$100~fb$^{-1}$ over the whole range of Higgs masses.

The undetected partial width can be constrained to be less than
$15-55\%$ of the total fitted Higgs width for 2$\times$30~fb$^{-1}$
and $15-30\%$ for 2$\times$300 + 2$\times$100~fb$^{-1}$, at the
$1\sigma$ level.  This undetected partial width is most tightly
constrained for Higgs masses above 160~GeV.

\smallskip

The resulting precisions on the Higgs boson couplings squared are
shown in Fig.~\ref{fig:fit} as a function of Higgs mass for the same
luminosity scenarios, 2$\times$30~fb$^{-1}$ and 2$\times$300 +
2$\times$100~fb$^{-1}$, and SM rates observed.  For the latter case,
typical accuracies range between 20 and $40\%$ for Higgs masses below
150~GeV.  Above $W$-pair threshold the measurement of the
then-dominant $H\to WW,ZZ$ partial widths improves to the $10\%$
level.  The case of 2$\times$300~fb$^{-1}$ yields only small
improvements over the right-hand panel in Fig.~\ref{fig:fit}, except
in the case of $g^2(H,\tau)$ which shows moderate improvement.
However, since this happens for Higgs masses below $\sim 140$~GeV,
this effect can be relatively important in the case of MSSM analyses,
see Sec.~\ref{sec:mssm_specific}.  This can be understood because the
$H\to\tau\tau$ decay is measured only in WBF, and $g(H,\tau)$ does not
have a large effect on the Higgs total width or loop-induced
couplings.

%It is interesting to note that if the ``LEP Higgs'' at 116 GeV were
%realized~\cite{Barate:2003sz}, the accuracy in the low luminosity case 
%is rather bad. In the higher luminosity
%case, on the other hand, the region around 116 GeV is not worse than
%the rest of the Higgs mass range allowed in the MSSM.

%%%%%%%%%%%%%%%%%%%%%%%%%%%%%%%%%%%%%%%%%%%%%%%%%%%%%%%%%%%%%%%%%%%%%%%%%%%%%%%
%%%%%%%%%%%%%%%%%%%%%%%%%%%%%%%%%%%%%%%%%%%%%%%%%%%%%%%%%%%%%%%%%%%%%%%%%%%%%%%

\subsection{Dominant systematic uncertainties}

The results shown in Fig.~\ref{fig:fit} reflect present understanding
of detector effects and systematic errors (see also the Appendix). One
should note that improved selection and higher acceptance will
decrease the statistical errors.  At least as important is work on the
reduction of systematic errors. In Fig.~\ref{fig:fit}, the thin lines
show expectations with vanishingly small systematics: systematic
errors contribute up to half the total error, especially at high
luminosity.

For a Higgs boson mass below 140 GeV the main contribution to the
systematic uncertainty is the background normalization from sidebands.
The largest contribution is from $H\to b\bar{b}$. For this channel the
signal to background ratio is between 1:4 and 1:10. For the background
normalization we assume a systematic error of
$10\%$~\cite{ATL-PHYS-2003-024}. This leads to a huge total systematic
error on the measurement of $\Gamma_b$, which is the main contribution
to the total width $\Gamma$ (the BR($H\to b\bar{b}$) is between $80\%$
and $30\%$). But a measurement of absolute couplings needs $\Gamma$ as
input (see discussion in the introduction to this section), so all
measurements of couplings share the large systematic uncertainty on
$H\to b\bar{b}$.

For a Higgs boson mass above 150~GeV there are two dominant
contributions to the systematic error:
\begin{itemize}
\item the background normalizations in GF, WBF and $t\bar{t}H$
  (systematic error $5\%$ to $15\%$)
\item the QCD uncertainty in the cross section calculations for GF
  ($20\%$) and $t\bar{t}H$ ($15\%$) from given Higgs boson couplings.
\end{itemize}
This is especially evident in the measurement of the top coupling
based on the $t\bar{t}H$ channel. Here the systematic uncertainties
contribute half of the total error.

%%%%%%%%%%%%%%%%%%%%%%%%%%%%%%%%%%%%%%%%%%%%%%%%%%%%%%%%%%%%%%%%%%%%%%%%%%%%%%%
%%%%%%%%%%%%%%%%%%%%%%%%%%%%%%%%%%%%%%%%%%%%%%%%%%%%%%%%%%%%%%%%%%%%%%%%%%%%%%%

\subsection{Additional constraints: SU(2) and SM loops}

The theoretical constraints used so far have been very moderate.  If,
in addition to the requirement that $g^2(H,W)<1.05 \cdot g^2(H,W,{\rm
  SM})$ and $g^2(H,Z)<1.05 \cdot g^2(H,Z,{\rm SM})$, we assume that no
new non-SM particles run in the loops for $H\to\gamma\gamma$ and
$gg\to H$ (which is approximately fulfilled for the MSSM with a not
too-light spectrum), the precision of the coupling measurements
improves only slightly, with the only noticeable improvement for Higgs
masses below 120~GeV.

%%%%%%%%%%%%%%%%%%%%%%%%% F I G U R E %%%%%%%%%%%%%%%%%%%%%%%%%%%%%%%%%%%%%%%%%
\begin{figure}[htb]
\begin{center}
\resizebox{\textwidth}{!}{
\includegraphics{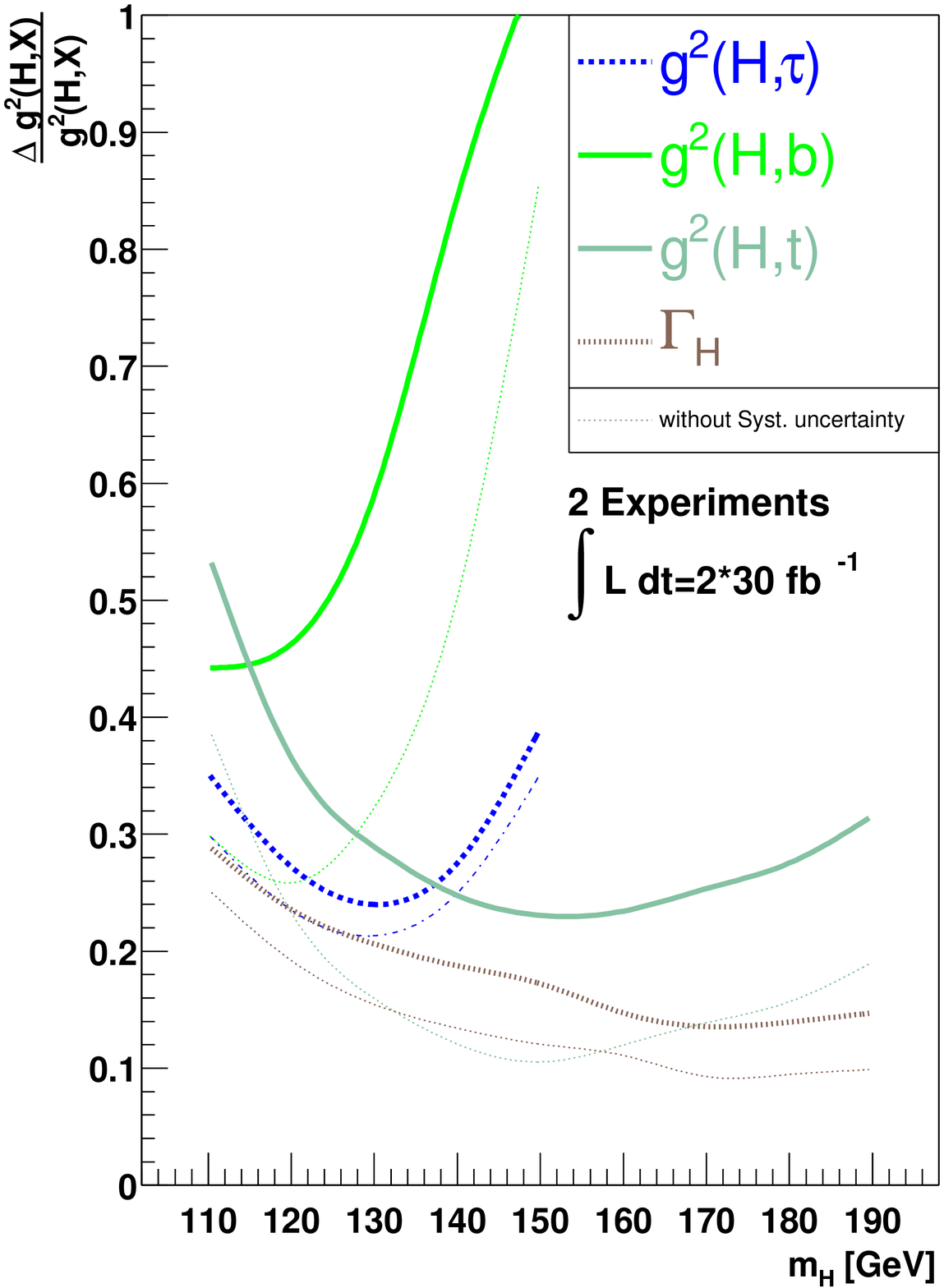}
\includegraphics{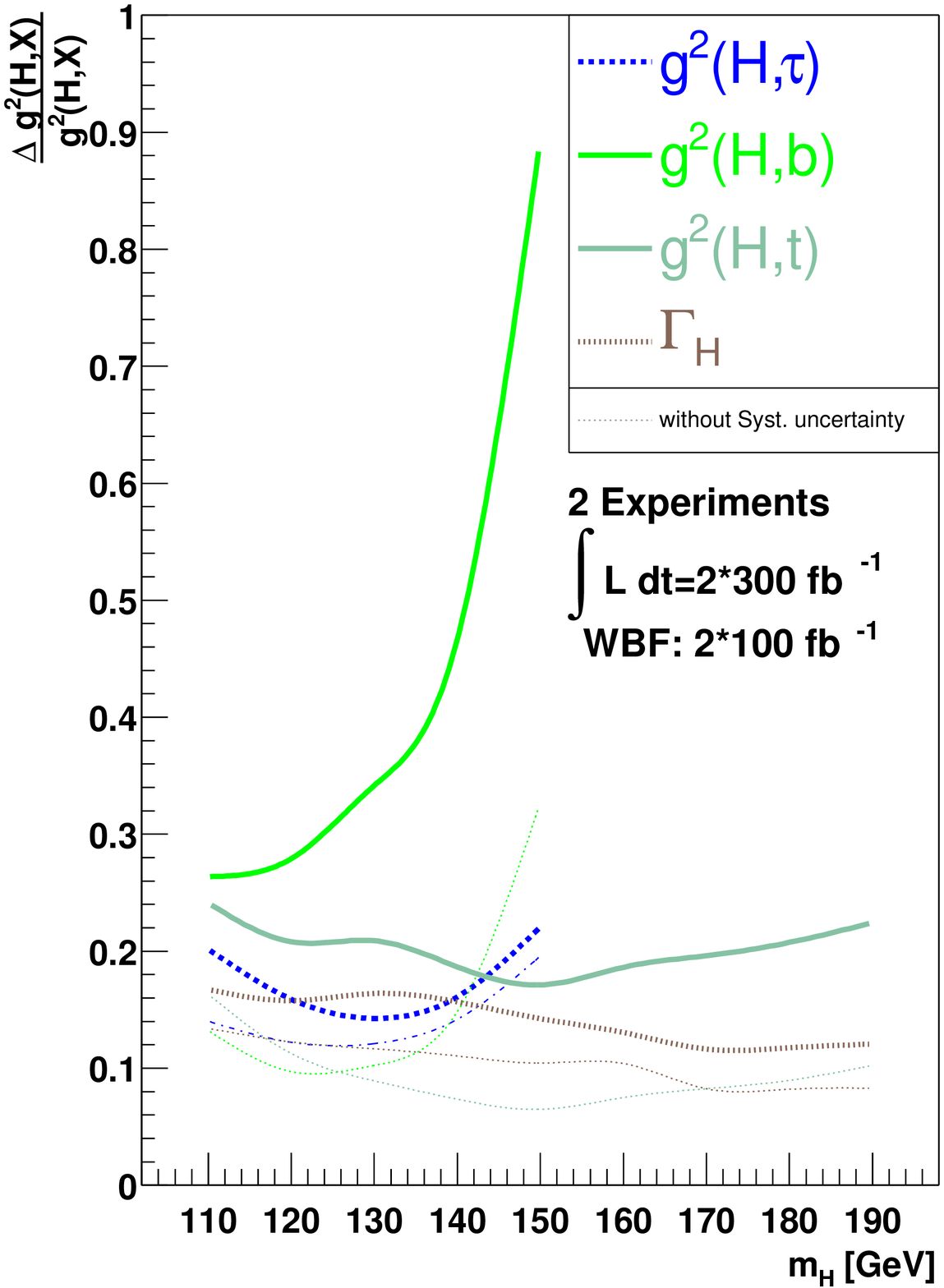}
}
\caption{As in Fig.~\ref{fig:fit}, but with more restrictive assumptions.
  Here we assume that $g^2(H,W)=g^2(H,W,{\rm SM})\pm 5\%$ and
  $g^2(H,W)/g^2(H,Z)=g^2(H,W,{\rm SM})/g^2(H,Z,{\rm SM})\pm 1\%$.  We
  also assume that no new particles run in the loops for
  $H\to\gamma\gamma$ and $gg\to H$, so that these couplings are fixed
  in terms of the couplings of the SM particles in the loops.  As in
  Fig.~\ref{fig:fit}, additional decays of the Higgs boson are fitted
  with a partial width for undetected decays (not shown).}
\label{fig:SU2fit}
\end{center}
\end{figure}
%%%%%%%%%%%%%%%%%%%%%%%%% F I G U R E %%%%%%%%%%%%%%%%%%%%%%%%%%%%%%%%%%%%%%%%%

Another small improvement comes about by restricting the $W$ and $Z$
couplings to their SM ratio.  Within the multi-Higgs-doublet models
considered throughout, SU(2) symmetry relates these two couplings.
It thus is natural to forgo an independent measurement of their ratio
and to rather assume that
\begin{equation}
\label{eq:su2}
g^2(H,W)/g^2(H,Z) = g^2(H,W,{\rm SM})/g^2(H,Z,{\rm SM}) \pm 1\%\;.
\end{equation}
Within the MSSM, this coupling ratio is indeed very close to its SM
value.

Over most of the MSSM parameter space even the individual $hVV$
couplings will be close to their SM values since decoupling sets in
rapidly once the mass of the $CP$-odd Higgs boson starts to become
large, $M_A\gsim 200$~GeV. This motivates a fit where in addition to
Eq.~\ref{eq:su2} we assume
\begin{equation}
g^2(H,W) = g^2(H,W,{\rm SM}) \pm 5\%\;.
\end{equation}
We again assume that no new non-SM particles run in the loops for
$H\to\gamma\gamma$ and $gg\to H$.  However, we fit additional Higgs
boson decays with a partial width for undetected decays.  The
constraints on this undetected partial width are essentially the same
as in our least-constrained fit, see Fig.~\ref{fig:Gnew}.  The
resulting parameter precisions are shown in Fig.~\ref{fig:SU2fit} and
reach $10-20\%$ over the entire intermediate Higgs mass range for the
2$\times$300 + 2$\times$100~fb$^{-1}$ luminosity scenarios. The only
exception is $g^2(H,b)$, which can be measured only to about $30\%$
for $m_H\lsim 130$~GeV.

Relaxing assumptions slightly, by allowing non-SM particles to
contribute to the $H\to\gamma\gamma$ partial width, has a noticeable
effect on the coupling determination only for $m_H\lsim 120$ GeV.  For
example, for the 2$\times$300 + 2$\times$100~fb$^{-1}$ luminosity
scenario, the precision on $g^2(H,\tau)$, $g^2(H,b)$ and the Higgs
total width at $m_H=110$ GeV jump to about $40\%$.

%%%%%%%%%%%%%%%%%%%%%%%%%%%%%%%%%%%%%%%%%%%%%%%%%%%%%%%%%%%%%%%%%%%%%%%%%%%%%%%
%%%%%%%%%%%%%%%%%%%%%%%%%%%%%%%%%%%%%%%%%%%%%%%%%%%%%%%%%%%%%%%%%%%%%%%%%%%%%%%

\section{Higgs couplings within the MSSM}
\label{sec:mssm_specific}

If the obtained Higgs boson couplings differ from the SM predictions,
one can investigate at what significance the SM can be excluded from
LHC measurements in the Higgs sector alone.  As a specific example of
physics beyond the SM, we consider the MSSM.

If supersymmetric partners of the SM particles were detected at the
LHC, this would of course rule out the SM.  It would nevertheless be
of interest in such a situation to directly verify the non-SM nature
of the Higgs sector.  Besides the possible detection of the additional
states of an extended Higgs sector, a precise measurement of the
couplings of the lightest (SM-like) Higgs boson will be crucial.

%A plausible scenario is that one or several Higgs bosons will be
%discovered at the LHC together with evidence for supersymmetry (SUSY)
%at the TeV scale.  Once SUSY had been confirmed, we would be led to
%analyze the Higgs sector in terms of a two Higgs doublet model with
%MSSM constraints.

For the sake of brevity let us assume that the pseudoscalar Higgs and
the charged Higgs are fairly heavy ($M_A\gsim 150$~GeV, and they may,
but need not, have been observed directly) so that they do not
interfere with the $h$~signal extraction. We furthermore assume that
only decays into SM particles are detected.  Then the light Higgs that
we consider here will have couplings to the $W$ and $Z$ which are
suppressed by the same factor $\sin(\alpha-\beta)$ compared to SM
strength, and Higgs couplings to fermions in addition depend on
$\tan\beta=v_2/v_1$ and $\Delta_b$~\cite{Hall:1994gn}, which
incorporates non-universal loop corrections to the $h\bar{b}b$
coupling.  A fit of the Higgs couplings can then be performed in terms
of this reduced parameter set.  Obviously this analysis falls within
the $g_V\le g_V^{\rm SM}$ analysis described in the previous section.
Upper bounds on the expected measurement errors for MSSM partial
widths can hence be derived from Fig.~\ref{fig:fit}, while
Fig.~\ref{fig:SU2fit} gives an estimate of errors which can be
expected for $M_A\gsim 200$~GeV, for which the Higgs couplings to $W$
and $Z$ bosons have sufficiently approached their SM values.

\begin{figure}[htb!]
\begin{center}
\resizebox{\textwidth}{!}{
\rotatebox{270}{\includegraphics[50,50][555,590]{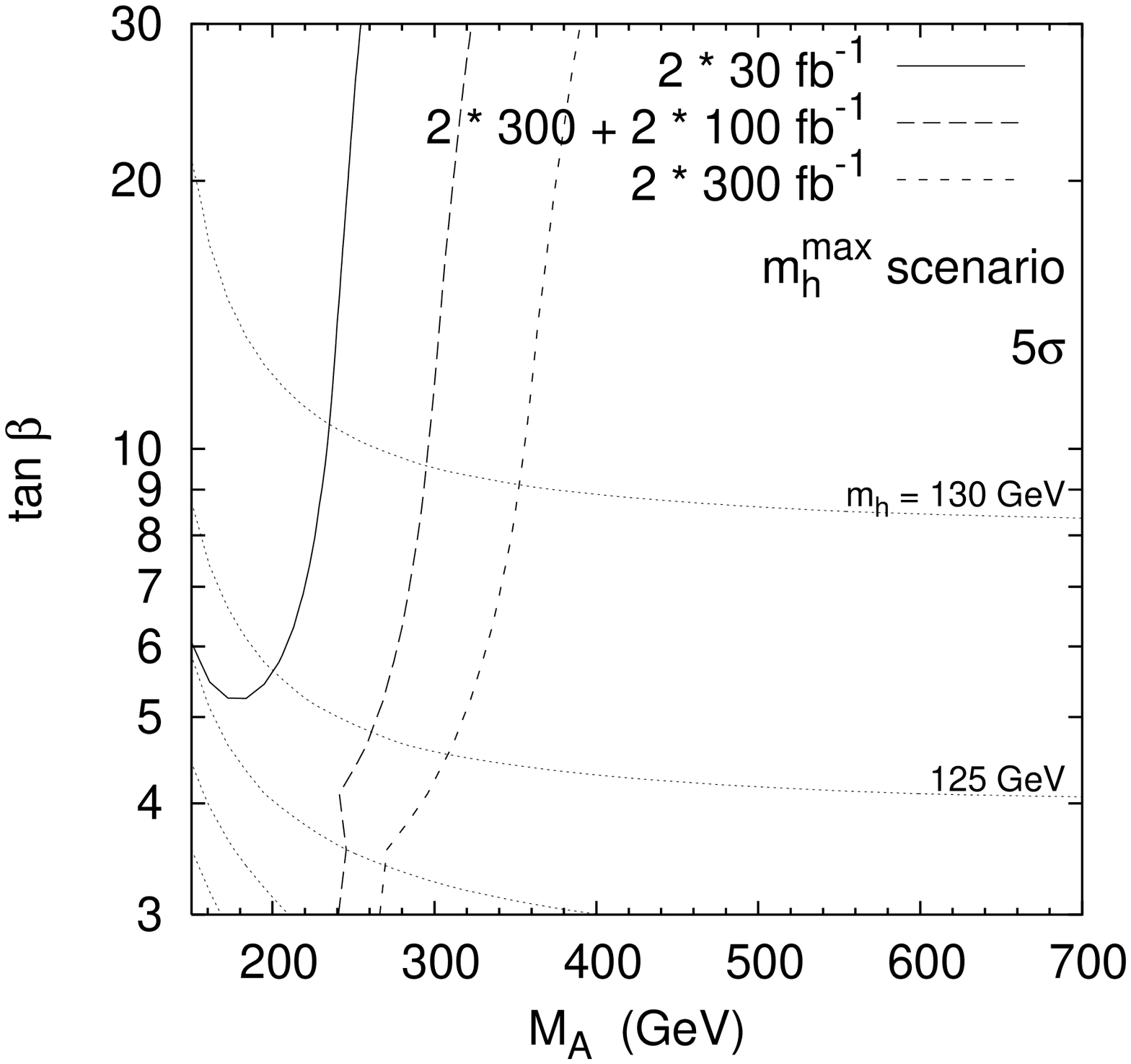}}
\rotatebox{270}{\includegraphics[50,50][555,590]{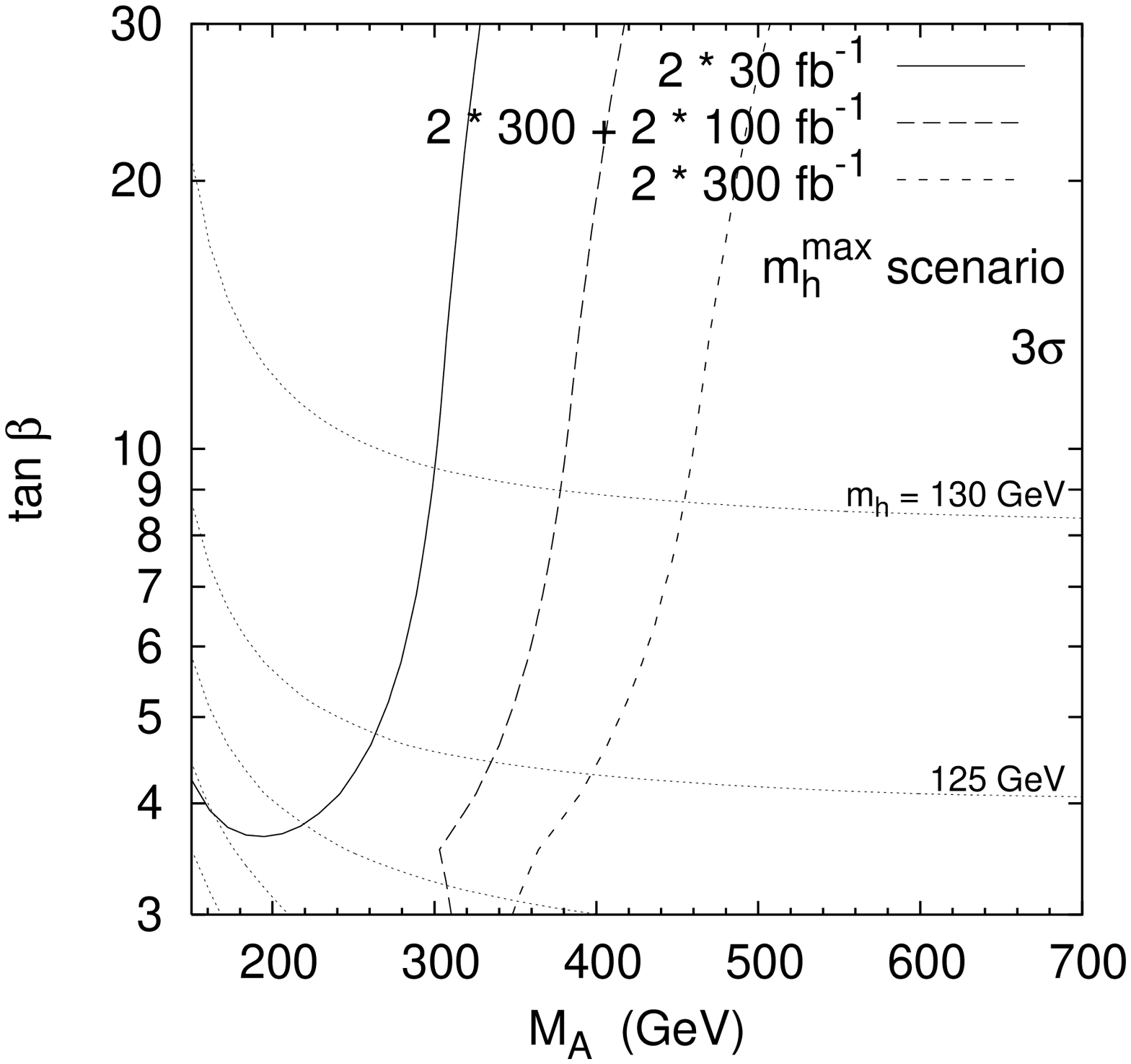}}
}
\caption{Fit within the MSSM $m_h^{\rm max}$ scenario in the 
  $M_A$--$\tan\beta$ plane for three luminosity scenarios.  The two
  panels show the region (to the left of the curves) which would yield
  a $\geq 5\sigma$ ($\Delta\chi^2\geq 25$) or $\geq 3\sigma$
  ($\Delta\chi^2\geq 9$) discrepancy from the SM.  The
  mostly-horizontal dotted lines are contours of $m_h$ in steps of
  5~GeV.  }
\label{fig:mssm_contour}
\end{center}
%\vspace{1cm}
\end{figure}

A quantitative, global measure of how well the LHC can distinguish the
SM from a specific MSSM scenario is provided by a $\chi^2$-analysis of
the deviations expected in that case.  As a first example we consider
the $m_h^{\rm max}$ scenario of Ref.~\cite{Carena:2002qg}.  We
calculate the mass and branching fractions of the MSSM Higgs boson
using HDECAY3.0 \cite{Djouadi:1998yw}, using the FeynHiggsFast1.2.2
\cite{Heinemeyer:1999be,Heinemeyer:2000nz} option to compute the MSSM
Higgs masses and couplings. Assuming that, for a given $M_A$ and
$\tan\beta$, the corresponding SUSY model is realized in nature, we
may ask at what significance the SM would be ruled out from
$h$~measurements alone. We examine MSSM points only for $M_A>150$~GeV,
where the narrow width approxmation is still valid.  The resulting
contours are shown in Fig.~\ref{fig:mssm_contour} for the three
luminosity assumptions defined in Sec.~\ref{subsec:fitproc}.  In the
areas to the left of the contours the SM can be rejected with more
than $5\sigma$ or $3\sigma$ significance, respectively.

%%%%%%%%%%%%%%%%%%%%%%%%% F I G U R E %%%%%%%%%%%%%%%%%%%%%%%%%%%%%%%%%%%%%%%%%
%%%%%%%%%%%%%%%%%%%%%%%%% F I G U R E %%%%%%%%%%%%%%%%%%%%%%%%%%%%%%%%%%%%%%%%%

The $\chi^2$ definition in Fig.~\ref{fig:mssm_contour} assumes the
same systematic errors as our analysis in Sec.~\ref{sec:nomodel}.
Event rates and resulting statistical errors, however, are those
expected for the MSSM.

\smallskip

The source of the MSSM analysis sensitivity can be understood as
follows.  In the $m_h^{\rm max}$ scenario for $M_A\gsim 200$ GeV, the
couplings of $h$ to SM particles all essentially obtain their SM
values except for the $hbb$ and $h\tau\tau$ couplings, due to the
slower decoupling behavior of the latter.  In this scenario the SUSY
threshold corrections to the $b$ mass are also quite small, so that
the ratio of the $hbb$ and $h\tau\tau$ couplings essentially takes its
SM value.  The $h\to b\bar{b}$ decay mode dominates the Higgs total
width in this scenario.  The pattern of Higgs coupling deviations can
then be summarized as follows: all the Higgs production cross sections
considered in our study are SM-like; the partial widths into
$b\bar{b}$ and $\tau\tau$ are equally enhanced (but with SM-like BRs
since the total width is dominated by $b\bar{b}$ and $\tau\tau$
decays).  This results in a larger total width for the Higgs boson.
The branching ratios into all other final states ($WW^{*}$, $ZZ^{*}$,
$\gamma\gamma$) are smaller than in the SM, reflecting this total
width enhancement.

We focus on the WBF channels, which have the strongest impact on the
MSSM fit.  If systematic errors could be neglected, the well-measured
WBF $qqH\to qqWW^{*}$ channel would give the best sensitivity to the
discrepancy from the SM, since it is sensitive to the Higgs branching
ratio into $WW^{*}$.  The less-well-measured WBF $qqH\to
qqZZ^{*},qq\gamma\gamma$ channels could be added to increase the
statistics.  However, the systematic uncertainties on the luminosity
($5\%$), WBF cross section ($4\%$), and forward tagging/veto jets
($5\%$) hide this sensitivity to the Higgs coupling deviations.  These
systematic uncertainties can be brought under control by including the
WBF $qqH\to qq\tau\tau$ channel in the fit.  While this channel by
itself provides no sensitivity to the deviation from the SM (because
the WBF cross section and the branching ratio to taus are both
SM-like), it serves to normalize out the systematic uncertainties.  To
a first approximation, this can be thought of as taking the ratio of
the WBF rates with Higgs decays to $WW^{*}$ ($ZZ^{*}$, $\gamma\gamma$)
versus $\tau\tau$, in which the aforementioned systematic
uncertainties cancel.  The $\chi^2$ fit of the rates in these channels
offers a slight improvement over the ratio method because the
systematic uncertainties are somewhat better under control.

%%%%%%%%%%%%%%%%%%%%%%%%% F I G U R E %%%%%%%%%%%%%%%%%%%%%%%%%%%%%%%%%%%%%%%%%
\begin{figure}[htb!]
\begin{center}
\resizebox{10cm}{!}{
\rotatebox{270}{\includegraphics[50,50][555,590]{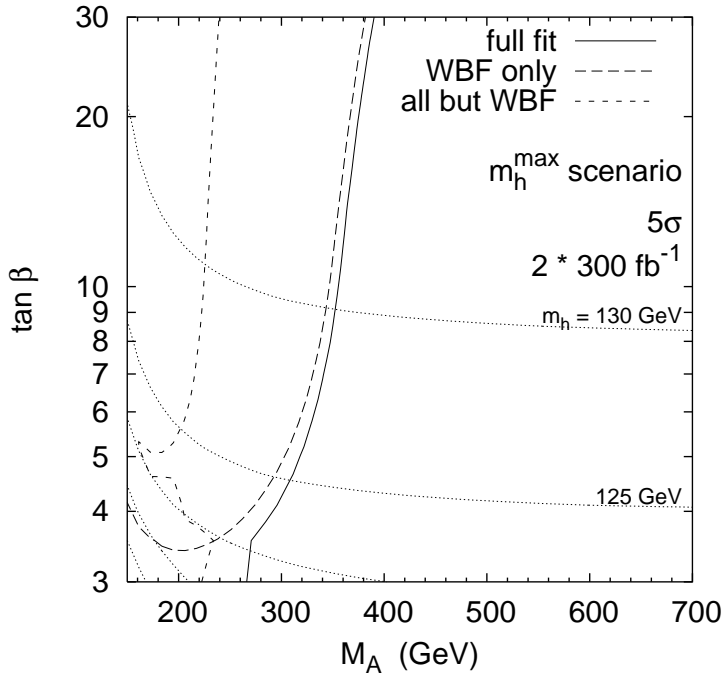}}
}
\caption{Fit within the MSSM $m_h^{\rm max}$ scenario in the 
  $M_A$--$\tan\beta$ plane for the 2$\times$300~fb$^{-1}$ luminosity
  scenario.  The $5\sigma$ lines are shown using WBF channels only
  (long--dashed), all channels except WBF (short--dashed), and the
  full fit (solid line).  }
\label{fig:mssm_sensitivity}
\end{center}
\end{figure}
%%%%%%%%%%%%%%%%%%%%%%%%% F I G U R E %%%%%%%%%%%%%%%%%%%%%%%%%%%%%%%%%%%%%%%%%

In Fig.~\ref{fig:mssm_sensitivity} we analyze the impact of the
different channels included in our analysis. Within the $m_h^{\rm
  max}$ scenario we show the $5\sigma$ contours in the
2$\times$300~fb$^{-1}$ luminosity scenario in the $M_A$--$\tan\beta$
plane. We plot separately the fit using only the WBF channels
(long--dashed), using all channels except WBF (short--dashed), and the
full fit (solid).  It becomes obvious that the WBF channels have the
strongest impact.  Therefore it will be very helpful for this kind of
analysis if the WBF channels can be fully exploited at high luminosity
running.

%%%%%%%%%%%%%%%%%%%%%%%%% F I G U R E %%%%%%%%%%%%%%%%%%%%%%%%%%%%%%%%%%%%%%%%%
\begin{figure}[htb!]
\begin{center}
\resizebox{\textwidth}{!}{
\rotatebox{270}{\includegraphics[50,50][555,590]{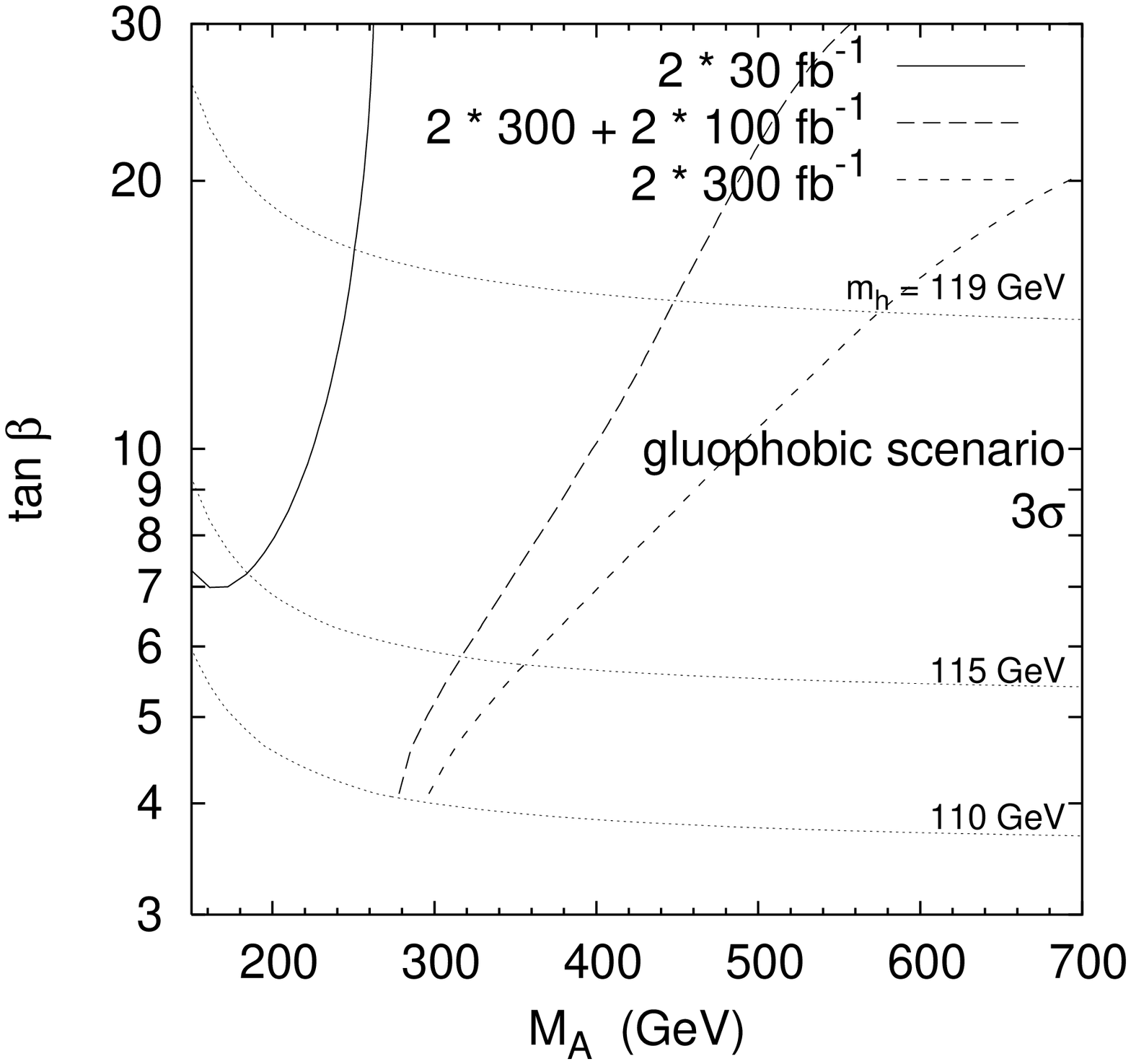}}
\rotatebox{270}{\includegraphics[50,50][555,590]{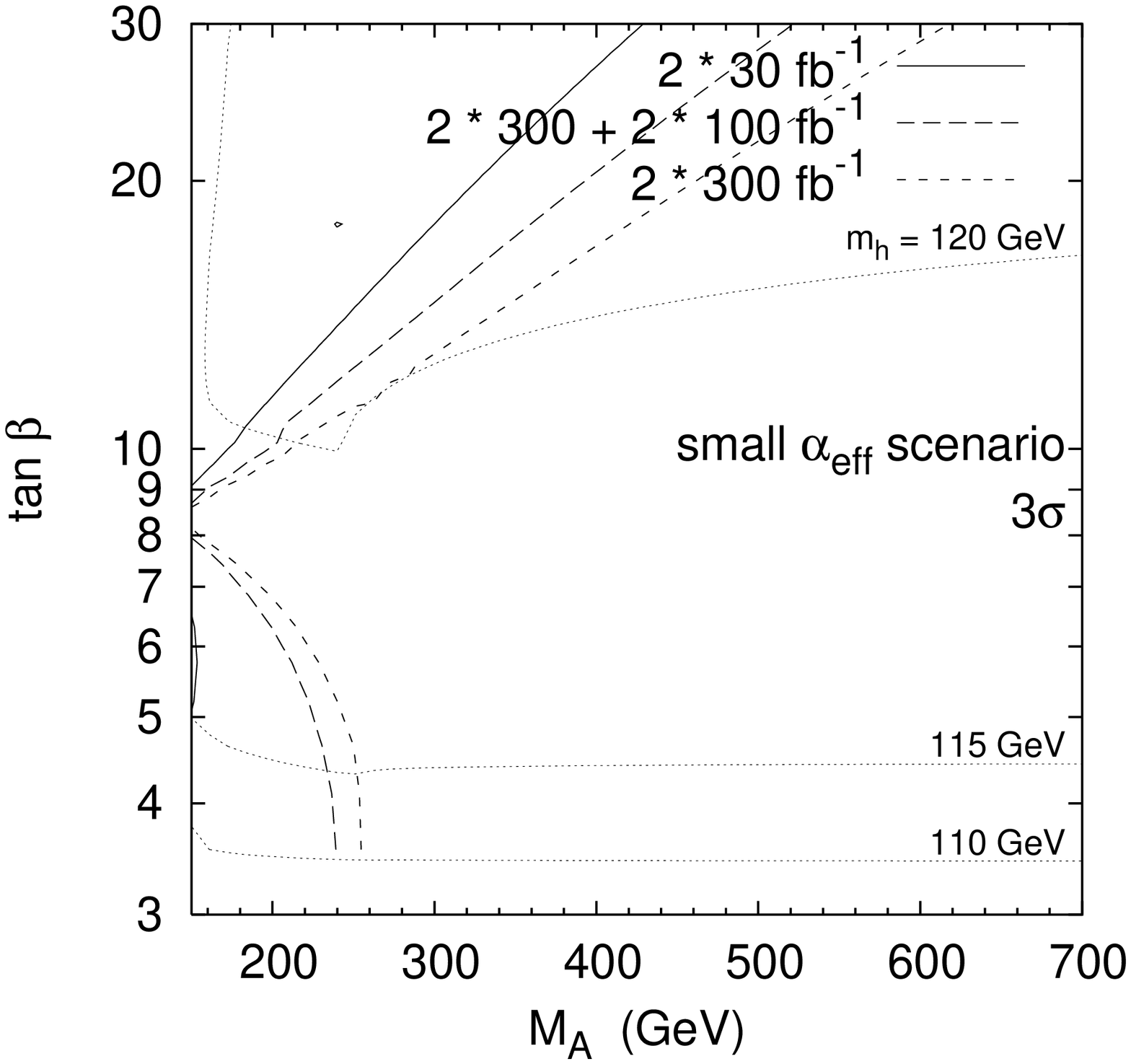}}
}
\caption{Fit within the MSSM ``gluophobic Higgs'' (left) and the
  ``small $\alpha_{\rm eff}$'' (right) scenario (see text) in the
  $M_A$--$\tan\beta$ plane for three luminosity scenarios.  The two
  panels show the region (to the left of the curves) in which a $\geq
  3\sigma$ ($\Delta\chi^2\geq 9$) discrepancy from the SM would be
  observed.  The mostly-horizontal dotted lines are contours of $m_h$,
  where the masses are indicated.}
\label{fig:mssm_contour2}
\end{center}
\end{figure}
%%%%%%%%%%%%%%%%%%%%%%%%% F I G U R E %%%%%%%%%%%%%%%%%%%%%%%%%%%%%%%%%%%%%%%%%

Finally we would like to emphasize that the contours appear
significantly different for different SUSY scenarios: other SUSY
parameters can have a large effect on the relation of $m_h$, $M_A$,
$\tan\beta$ and Higgs couplings. This is shown in
Fig.~\ref{fig:mssm_contour2}, where the $3\sigma$ significance curves
are shown in the ``gluophobic Higgs'' (left) and the ``small
$\alpha_{\rm eff}$'' (right) scenarios~\cite{Carena:2002qg}. In the
gluophobic Higgs scenario the $ggh$ coupling is strongly suppressed
over the whole $M_A$--$\tan\beta$ plane due to additional
contributions from scalar top loops.  We note that for this scenario
the $5\sigma$ curves (not shown) move significantly to the left.
Within the small~$\alpha_{\rm eff}$ scenario loop corrections can
suppress the $hbb$ and $h\tau\tau$ couplings for moderate values of
$M_A$ and large values of $\tan\beta$.  Both effects yield a strong
deviation from the corresponding $3\sigma$ curves obtained in the
$m_h^{\rm max}$ scenario.

It should be noted that the shown sensitivity to $M_A$ cannot directly
be translated into indirect bounds on $M_A$.  To establish realistic
bounds, a careful analysis of the experimental errors arising from the
incomplete knowledge of the spectrum of supersymmetric particles and
of the theoretical uncertainties from unknown higher-order corrections
is necessary.  However, we observe that the contours shift only
slightly if one instead uses SM rates to calculate the statistical
errors.

%It should also be kept in mind this the analysis of this section does
%not apply to non-minimal extensions of the SUSY Higgs sector, since
%they may exhibit strong deviations from the MSSM patterns.  Hence, an
%indirect determination of $M_A$ from observed deviations in light
%Higgs couplings is problematic without significant input from other
%direct SUSY observations at the LHC.

%%%%%%%%%%%%%%%%%%%%%%%%%%%%%%%%%%%%%%%%%%%%%%%%%%%%%%%%%%%%%%%%%%%%%%%%%%%%%%%
%%%%%%%%%%%%%%%%%%%%%%%%%%%%%%%%%%%%%%%%%%%%%%%%%%%%%%%%%%%%%%%%%%%%%%%%%%%%%%%

\section{Summary and outlook}
\label{sec:sum_out}

Measurements of the Higgs sector are expected to provide many
complementary signatures after several years of LHC running. Combining
these measurements allows one to extract information on Higgs partial
widths and Higgs couplings to fermions and gauge bosons. Because
significant contributions from unobservable channels cannot easily be
ruled out at the LHC, model-independent analyses produce large
correlations between extracted partial widths. A reduction of
correlations and hence smaller errors on particular couplings can be
achieved with only very weak theory assumptions applicable to quite
general Higgs sector scenarios.  In this paper we have analyzed the
constraints expected in generic multi-Higgs-doublet models, namely
that $HVV$ couplings cannot be larger than within the SM. Within such
models, the LHC can measure Higgs couplings to the top quark, tau
lepton, and $W$ and $Z$ bosons with accuracies in the $10-40\%$ range
once 300~fb$^{-1}$ of data have been collected.  If, on the other
hand, the SLHC will be realized, one could hope for significant
improvements over the results presented here. This applies in
particular for the bottom Yukawa coupling determination.

Within the MSSM, significant deviations in the Higgs sector should be
observable at the LHC, provided that the charged and the pseudoscalar
Higgs masses are not too heavy, i.e., that decoupling is not
completely realized. For example, within the $m_h^{\rm max}$ scenario
and with 300~fb$^{-1}$ of data, the LHC can distinguish the MSSM and
the SM at the $3\sigma$ level up to $M_A\simeq 450$~GeV and with
$5\sigma$ significance up to $M_A\simeq 350$~GeV with the Higgs data
alone.  The LHC will thus provide a surprisingly sensitive first look
at the Higgs sector, even though it cannot match the precision and
model-independence of analyses which are expected for a linear
$e^+e^-$
collider~\cite{Aguilar-Saavedra:2001rg,Abe:2001np,Abe:2001gc}.

\bigskip

So far we have investigated the situation where no important channel
suffers substantial suppression. However, it might be (within
Supersymmetry or another extension of the SM) that the WBF channels
are degraded, or that the Higgs decays more strongly to unobservable
$c\bar{c}$ or $gg$ final states. Other decays like
$h\to\tilde\chi_1^0\tilde\chi_1^0$~\cite{Eboli:2000ze} or
$h\to\mu^+\mu^-$~\cite{hmumu} may be detectable, or upper bounds may
be put on their partial widths. Channels with low statistics might be
absent. Finally, the mass measurement might be less precise due to a
suppression of $H\to\gamma\gamma$, thus weakening the Higgs mass
constraint. These scenarios are beyond the scope of this paper and
will be discussed in a forthcoming publication.

%%%%%%%%%%%%%%%%%%%%%%%%%%%%%%%%%%%%%%%%%%%%%%%%%%%%%%%%%%%%%%%%%%%%%%%%%%%%%%%
%%%%%%%%%%%%%%%%%%%%%%%%%%%%%%%%%%%%%%%%%%%%%%%%%%%%%%%%%%%%%%%%%%%%%%%%%%%%%%%
 
\section*{Acknowledgments}

We would like to thank M.~Carena for useful discussions during early
stages of this work.  H.L. and D.Z. were supported in part by the
U.S.~Department of Energy under grant DE-FG02-95ER40896 and in part by
the Wisconsin Alumni Research Foundation.  This work was also
supported by the European Community's Human Potential Programme under
contract HPRN-CT-2000-00149 Physics at Colliders.  S.H. thanks the
DESY theory division for kind hospitality in the final stages of this
work.

%%%%%%%%%%%%%%%%%%%%%%%%%%%%%%%%%%%%%%%%%%%%%%%%%%%%%%%%%%%%%%%%%%%%%%%%%%%%%%%
%%%%%%%%%%%%%%%%%%%%%%%%%%%%%%%%%%%%%%%%%%%%%%%%%%%%%%%%%%%%%%%%%%%%%%%%%%%%%%%

\begin{appendix}
\section*{Appendix: systematic uncertainies}

The systematic errors include uncertainties on luminosity and detector
effects which are summarized in Tab.~\ref{Tabl:SystDetector}.  All
these numbers are estimates.  More definite numbers will be known only
once the LHC experiments are running.

\begin{table}[h]
\begin{center}
\begin{tabular}{|l@{\hspace{4ex}}l@{\hspace{6ex}}l|} \hline
  $L$                   & $5\%$ & Measurement of luminosity             \\ \hline
  $\epsilon_D$          & $2\%$ & Detector efficiency                   \\ \hline
  $\epsilon_L$          & $2\%$ & Lepton reconstruction efficiency      \\ \hline
  $\epsilon_\gamma$     & $2\%$ & Photon reconstruction efficiency      \\ \hline
  $\epsilon_b$          & $3\%$ & $b$-tagging efficiency                \\ \hline
  $\epsilon_\tau$       & $3\%$ & hadronic $\tau$-tagging efficiency    \\ \hline
  $\epsilon_{\rm{Tag}}$ & $5\%$ & WBF tag-jets / jet-veto efficiency    \\ \hline
  $\epsilon_{\rm{Iso}}$ & $3\%$ & Lepton isolation ($H\to ZZ\to 4\ell$) \\ \hline
\end{tabular}
\caption{\label{Tabl:SystDetector}
  Estimated systematic uncertainties on luminosity and detector
  effects, see e.g. Ref.~\cite{ATL-PHYS-2003-030}.}
\end{center}
\end{table}

The systematic background normalization uncertainties of the
individual channels are split into two components, shown in the second
and third column of Tab.~\ref{Tabl:SystBackground}.  The first part is
the uncertainty on the shape of the background derived from
extrapolating a perfectly measured sideband into the signal region.
The second part is needed to estimate the statistical error on the
measurement of the sideband itself.  We used this manner of estimating
the number of events in the sideband since actual numbers for
sidebands are not contained in the existing analyses.

\begin{table}[h]
\begin{center}
\begin{tabular}{|l@{\hspace{0.5ex}}|r|r|}
  \hline
  \multicolumn{1}{|c|}{Decay}&
  \multicolumn{1}{c|}{Shape}& \multicolumn{1}{c|}{$N_N/N_B$} \\ \hline
  $H\to ZZ^{(*)}\to 4l$              &  $1\%$  &  5 \\ \hline
  $H\to WW^{(*)}\to\ell\nu\,\ell\nu$ &  $5\%$  &  1 \\ \hline
  $H\to\gamma\gamma$                 & $0.1\%$ & 10 \\ \hline
  $H\to\tau\tau $                    &  $5\%$  &  2 \\ \hline
  $H\to b\bar{b}$                    & $10\%$  &  1 \\ \hline
\end{tabular}
\caption{\label{Tabl:SystBackground}
  Estimated systematic uncertainties on background normalization.  The
  extrapolation of a measured sideband into the signal region results
  in an uncertainty on the extrapolated number of background events.
  This contribution is given in the second column as systematic error
  on the shape.  The third column shows the ratio $N_N/N_B$ where
  $N_N$ is the actual number of events in the sideband and $N_B$ is
  the expected number of background events in the signal region. The
  ratio is used to estimate the statistical error on the measurement
  of the sideband itself from the numbers $N_B$ given in the
  experimental studies.}
\end{center}
\end{table}

The uncertainties in Tab.~\ref{Tabl:QCD} summarize the theoretical QCD
and PDF uncertainties on Higgs boson production.  For the WBF channels
there is an additional $10\%$ (after applying the minijet veto)
contribution from $gg\to Hgg$~\cite{DelDuca:2001eu}, which has its own
theory uncertainty of a factor of 2.

\begin{table}[h]
\begin{center}
\begin{tabular}{|l|l|} \hline
  GF          &  $20\%$ \\ \hline
  $t\bar{t}H$ &  $15\%$ \\ \hline
  $WH$        &   $7\%$ \\ \hline
  $ZH$        &   $7\%$ \\ \hline
  WBF         &   $4\%$ \\ \hline
  $gg\to Hgg$ & $100\%$ \\ \hline
\end{tabular}
\caption{\label{Tabl:QCD}
  Theoretical QCD and PDF uncertainties on the various Higgs boson
  production channels. The channel $gg\to Hgg$ was added to all WBF
  analyses at $10\%$ of the WBF rate with an uncertainty of a factor
  2.}
\end{center}
\end{table}
\end{appendix}

\vspace{-1cm}

%%%%%%%%%%%%%%%%%%%%%%%%%%%%%%%%%%%%%%%%%%%%%%%%%%%%%%%%%%%%%%%%%%%%%%%%%%%%%%%
%%%%%%%%%%%%%%%%%%%%%%%%%%%%%%%%%%%%%%%%%%%%%%%%%%%%%%%%%%%%%%%%%%%%%%%%%%%%%%%

%%%%%%%%%%%%%%%%%%%%%%%%%%%%%%%%%%%%%%%%%%%%%%%%%%%%%%%%%%%%%%%%%%%%%%%%%%%%%%%
%%%%%%%%%%%%%%%%%%%%%%%%%%%%%%%%%%%%%%%%%%%%%%%%%%%%%%%%%%%%%%%%%%%%%%%%%%%%%%%

\end{document}